\documentclass[12pt]{iopart}

\usepackage{url}
\usepackage{graphicx}
\usepackage{subfigure}
\begin{document}

\title[Visual Analytics for Star Formation]{Vialactea Visual Analytics tool for Star Formation studies of the Galactic Plane}

\author{F. Vitello$^{1}$, E. Sciacca$^{1}$, U. Becciani$^{1}$, A. Costa$^{1}$, M. Bandieramonte$^{2}$, M. Benedettini$^{3}$, M. Brescia$^{5}$, R. Butora$^{4}$, S. Cavuoti$^{6}$, A.M. Di Giorgio$^{3}$, D. Elia$^{3}$, S. J. Liu$^{3}$, S. Molinari$^{3}$, M. Molinaro$^{4}$, G. Riccio$^{5}$, E. Schisano$^{3}$ and R. Smareglia$^{4}$}
\address{$^1$INAF-Astrophysical Observatory of Catania, Catania, Italy}
\address{$^2$CERN, 385 Route de Meyrin 1217 Meyrin Suisse}
\address{$^3$INAF-Institute for Astrophysics and Space Planetology, Rome, Italy}
\address{$^4$INAF-Astronomical Observatory of Trieste, Trieste, Italy}
\address{$^5$ INAF-Astronomical Observatory of Capodimonte, Napoli, Italy}
\address{$^6$ University of Napoli Federico II - Dept. of Physics E. Pancini, Napoli, Italy}
\ead{fabio.vitello@oact.inaf.it}

\begin{abstract}
We present a visual analytics tool, based on the VisIVO suite, to exploit a combination of all new-generation surveys of the Galactic Plane to study the star formation process of the Milky Way. The tool has been developed within the VIALACTEA project, founded by the 7th Framework Programme of the European Union, that creates a common forum for the major new-generation surveys of the Milky Way Galactic Plane from the near infrared to the radio, both in thermal continuum and molecular lines. Massive volumes of data are produced by space missions and ground-based facilities and the ability to collect and store them is increasing at a higher pace than the ability to analyze them. This gap leads to new challenges in the analysis pipeline to discover information contained in the data. Visual analytics focuses on handling these massive, heterogeneous, and dynamic volumes of information accessing the data previously processed by data mining algorithms and advanced analysis techniques with highly interactive visual interfaces offering scientists the opportunity for in-depth understanding of massive, noisy, and high-dimensional data.
\end{abstract}

\maketitle

\section{INTRODUCTION}
Nowadays in Astrophysics data are produced at a very-high rate, and the quantity of data collected and stored is increasing at a faster rate than the ability to analyze them. Massive volumes of unstructured data, coming from space missions and ground-based facilities, covering the whole frequency spectrum, form continuous streams of terabytes of data that have to be recorded and analyzed. Future large infrastructures such as the Square Kilometre Array\cite{SKA} (SKA) are estimated to generate an exabyte a day of raw data, which could be compressed to around 10 petabytes.

Modern astronomical surveys provide not only image data but also catalogues of millions of objects (for example stars and galaxies), each of them with tens of associated parameters. The amount and complexity of this data is so high that it far exceeds the ability of humans to evaluate it all. By common data analysis techniques like knowledge discovery, astronomers can find new phenomena, relationships and useful knowledge about the universe. Given that data are affected by noise, a visual analytics approach can help separating relevant data from noise and help identifying unexpected phenomena inside the massive and dynamic data streams.

Astronomy visualization \cite{hassan2011scientific} encompasses topics such as optical and radio imaging, presentation of simulation results, multi-dimensional exploration of catalogues, and public outreach visuals. It is of fundamental importance in various stages of astronomical research: from the planning phase, through the observing processes or simulation runs, to quality control, qualitative knowledge discovery and quantitative analysis. 

Many tools are available to astrophysicists for visualization and data access. For example, TOPCAT\cite{TOPCAT} is an interactive graphical viewer and editor for astronomical tables such as object catalogues, designed to cope well with large tables. It offers a variety of ways to visualize and analyse the data, and extensive 2- and 3-d visualisation facilities. A number of options are provided for loading data from external sources, including Virtual Observatory (VO) services, thus providing a gateway to many remote archives of astronomical data. It can also interoperate with other desktop tools using VO Simple Application Messaging Protocol (SAMP)\cite{SAMP} such as Aladin\cite{Aladin} or SAOImage DS9\cite{SAOImage}. Aladin allows to visualize and manipulate digitized astronomical images or full surveys, superimpose entries from astronomical catalogues or databases, and interactively access related data and information from the Simbad\cite{Simbad} database, the VizieR\cite{VizieR} service and other archives for all known astronomical objects in the field. SAOImage DS9 is an astronomical imaging and data visualization application. It supports FITS images and binary tables, multiple frame buffers, region manipulation, and advanced features such as 2-D, 3-D and RGB frame buffers, mosaic images, tiling, blinking, geometric markers, colourmap manipulation, scaling, arbitrary zoom, cropping, rotation, pan, and a variety of coordinate systems.
VisIVO (Visualization Interface for the Virtual Observatory) \cite{sciacca2015integrated}, which is the core on which the presented visual analytics environment is based, offers a unique integrated ecosystem for visualization including: services for collaborative portals, mobile applications for visualization and data exploration and a number of key components such as workflow applications, analysis and data mining functionalities.

In recent years, the visual analytics is an emerging area  of scientific analysis. It offers an integral approach combining visualization, human factors and data analysis \cite{keim2008visual}. It integrates methodology from information analytics, geospatial analytics, and scientific analytics. Especially human factors (e.g., interaction, cognition, perception, collaboration, presentation, and dissemination) play a key role in the communication between human and computer, as well as in the decision-making process \cite{hilda2016review}. Visual analytics furthermore profits from methodologies developed in the fields of data management and knowledge representation, knowledge discovery and statistical analytics.

As an example, \textit{encube} \cite{vohl2016collaborative} provides a large scale comparative visual analytics framework tailored for use with large tiled-displays and advanced immersive environments like the CAVE2 (a modern hybrid 2D and 3D virtual reality environment). The framework is designed to harness the power of high-end visualisation environments for collaborative analysis of large subsets of data from radio surveys.

This paper presents the ViaLactea Visual Analytic tool (VLVA), an innovative environment for the study of star forming regions on our Galaxy. It allows an integrated analysis and exploitation of the combination of all new-generation surveys (from infrared to radio) of the Galactic Plane from space missions and ground-based facilities, using a novel data and science analysis paradigm based on 3D visual analytics and data mining frameworks. We will show the main features of VLVA  via three fundamental use cases related to i) the analysis of compact sources clumps and Spectral Energy Distributions (SEDs); ii) studies on diffuse bubble-like structures; and iii) filamentary structure analysis. 

\section{BACKGROUND}


The formation of stars and star clusters is by far the most important event that shapes the evolution and fate of galaxies. Let us take for example massive stars, which are responsible for the global ionization of the Interstellar Medium (ISM) (see e.g. \cite{churchwell2002ultra} and references therein). Their energetic stellar winds and supernova blast waves direct the dynamical evolution of the ISM, shaping its morphology, energetics and chemistry, and influencing the formation of subsequent generations of stars and planetary systems.

Stars form in Molecular Clouds, where about half of the mass in the ISM is stored. Due to the strong extinction of optical light by dust grains mixed with the gas, the birth of stars must be studied at infrared (IR) and longer wavelengths that can penetrate the clouds \cite{rieke1985interstellar,martin1990interstellar,fitzpatrick2009analysis}. The paradigm for the formation of solar-type stars via accretion from an envelope through a circumstellar disk predicts an evolution from cores to protostars, and finally pre-main sequence stars that is well matched with distinctive characteristics of their Spectral Energy Distribution (SED).

The empirical classification of the SED of low mass Young Stellar Object (YSOs) either based on detailed modeling \cite{robitaille2006interpreting} or on simple colour analysis e.g. \cite{gutermuth2009spitzer}, as well as the dust envelope temperature and the evolution of integrated parameters like the bolometric luminosity and the mass of the circumstellar envelope \cite{molinari2008evolution} have been used as powerful tools to constrain the evolutionary stage of YSOs. The formation of low-mass stars is however complicated by the fact that they are born for the most part in clusters together with more massive YSO. Higher-mass YSOs reach the conditions for H-burning faster than the time required to assemble them, so that winds and radiative acceleration will strongly influence their late accretion phases \cite{zinnecker2007toward} and that of the other cluster members.

In a traditional ``slow formation'' picture, diffuse ISM is accumulated by large-scale perturbations such as the passage of a spiral arm. Shielding by dust and surface reactions on grains promotes the transition from HI to HII, which in turn allows the formation of other molecules that cool the cloud. Gravity, mediated by the magnetic fields leads to unstable dense clumps with typical sizes between 0.1 and 1 pc, that can further differentiate into a multiplicity of cores that are typically dubbed protoclusters. Typically each core harbors a forming YSOs eventually leading to a single star or a close binary. Alternative more dynamical scenarios posit that Molecular Clouds are transient, short-lived structures created in the post-shock regions of converging large-scale flows \cite{hartmann2001rapid} or in turbulence driven compression \cite{padoan2002stellar} that produce layers and filaments where column density builds up shielding material from the Interstellar Radiation field. Cooling and gravitational instabilities will naturally arise \cite{heitsch2008fragmentation}, fragmenting the filaments under the effect of self gravity into chains of dense supercritical turbulent clumps that harbour massive protostars and protoclusters. Indeed, one of the most important and unexpected discoveries of the \textit{Herschel} Galactic photometric surveys \cite{molinari2010clouds} is the Galaxy-wide ubiquity of the filamentary nature of the cold, dense and potentially star-forming phase of the ISM. Star formation could also be triggered in clumps located at the interfaces between dense clouds and expanding HII regions or supernova remnants \cite{deharveng2008massive}.

The fact that the mass function in protoclusters has a similar shape to the Initial Mass Function in field stars suggests that the end-products of star formation are entirely defined as early as the clump formation stage. Therefore the detailed morphology of dense star forming structures, their mass and temperature distribution, and the timescales for their formation and for star formation inside them, hold the key to identify the dominant mechanism responsible for their formation.


The EU FP7 VIALACTEA project\footnote{EU FP7 VIALACTEA project web page: \url{http://vialactea.iaps.inaf.it/}} aims to exploit the combination of all new-generation surveys of the Galactic Plane to build and deliver a galaxy scale predictive model for star formation of the Milky Way. Usually the essential steps necessary to unveil the inner workings of the galaxy as a star formation engine are often carried out manually, and necessarily over a limited number of galactic sources or very restricted regions. Therefore scientists required new technological solutions able to deal with the growing data size and quantity coming from the available surveys to extract the meaningful informations contained in the data.  This resulted in an innovative framework~\cite{becciani2015advanced} based on advanced visual analytics techniques (presented in this paper) accessing a knowledge base based on Virtual Observatory (VO) data representation and retrieval standards that is enriched thanks to data mining and machine learning methodologies. 

A new set of complete and high spatial resolution Galactic Plane Surveys has been assembled and stored into the ViaLactea Knowledge Base (VLKB)~\cite{molinaro2016vialactea}.
The VLKB storage contains files in FITS format, from 2D images in the radio continuum to 3D FITS cubes containing radio velocity spectra at specific molecular lines and, also, a collection of 3D extinction maps.  Apart from the Herschel infrared Galactic Plane Survey (Hi-GAL)\cite{molinari2010hi}, the core resource from which most of the primary products are derived, all of the other resources were retrieved from already public repositories or released to the community retaining their private data policy. Alongside the data collections, a relational database completes the VLKB resource content in terms of knowledge derived from the data and contains information related to: 2D maps, filaments and bubbles, compact sources and radio cubes.
The available surveys can be classified into:
\begin{itemize}
\item Near-IR: UKIDSS\cite{UKIDSS} Galactic Plane Survey and VISTA VVV\cite{VISTA} Survey.
\item Mid-IR: The GLIMPSE\cite{GLIMPSE}, WISE\cite{WISE} and MIPSGAL\cite{MIPSGAL} Surveys.
\item Far-IR: \textit{Herschel} Hi-GAL\cite{HI-GAL} Survey.
\item Sub-millimetre continuum: ATLASGAL\cite{csengeri2016atlasgal} and JCMT\cite{JCMT} surveys.
\item Molecular and Atomic Line Surveys: the Galactic Ring Survey\cite{GRS} (GRS) and the International Galactic Plane Survey\cite{IGPS} (IGPS)
\item Radio continuum: the CORNISH\cite{CORNISH} and MAGPIS\cite{helfand2006magpis} Surveys.
\item Molecular Masers: the Methanol Multi-Beam survey\cite{MMB} (MMB).
\end{itemize}

\section{USE CASES}

In this section we show three exemplifying use cases where VLVA provides a key tool in all phases of scientific analysis thanks to its tight integration with the VLKB. In particular we will focus on the investigation of stars cluster formation and on the study of extended sources and filamentary structures. 

VLVA combines different types of visualization to perform the analysis exploring the correlation between different data, for example 2-D intensity images with 3-D molecular spectral cubes. The scientist is enabled to discover the link between different physical structures, from the extended filamentary-shaped structures to the most compact, dense sources precursors of new stars. 

\begin{figure*}
\includegraphics[width=\textwidth]{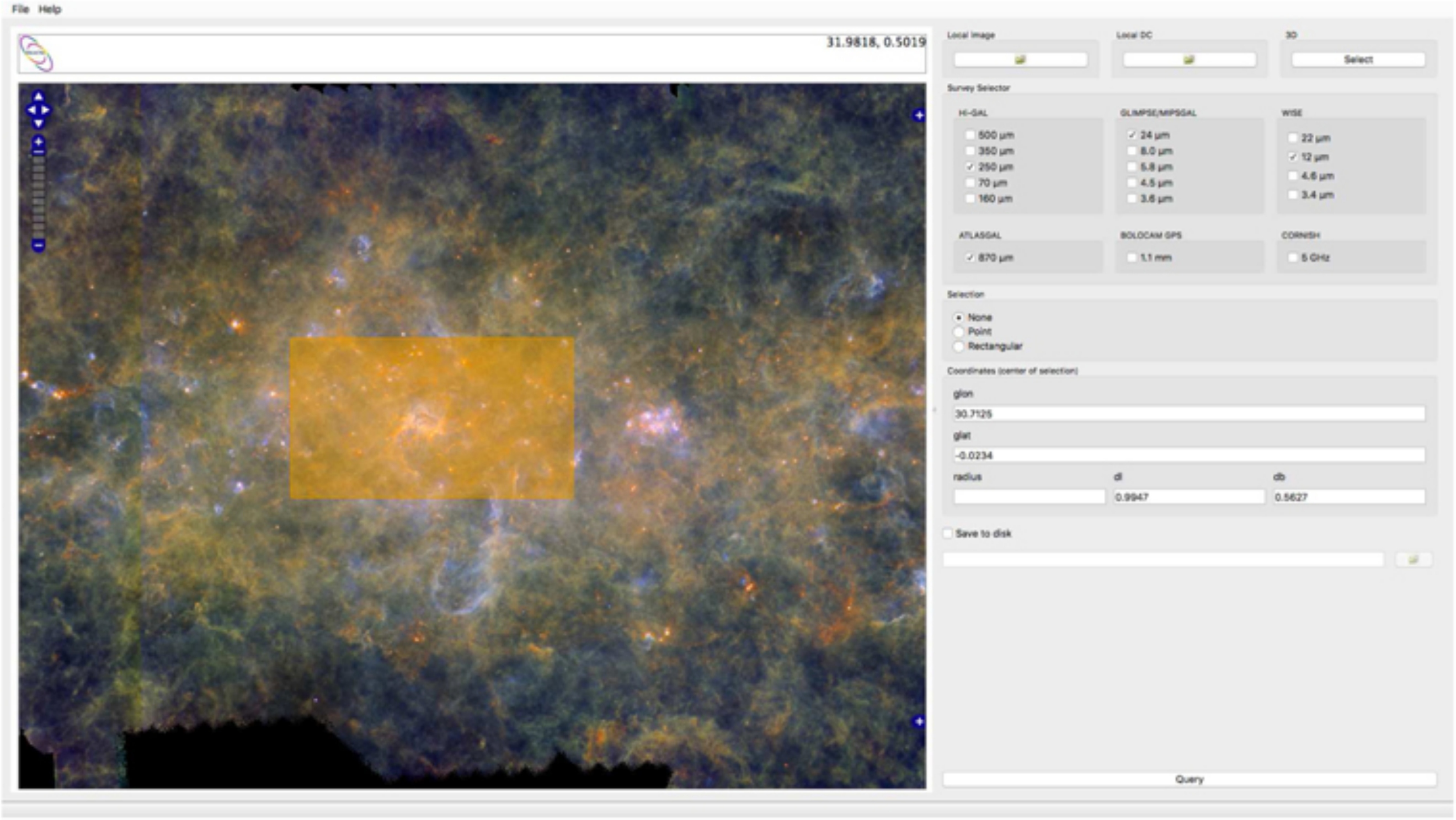}
\caption{Galactic Plane navigation on the  maps \textit{Herschel} Hi-GAL maps and rectangular region selection to start the analysis process.}
\label{fig1}
\end{figure*}

The implementation philosophy behind VLVA is to make transparent to the scientist the access to all informations without requiring technical skills to query the VLKB. This easy accessibility to the huge amount of data is offered since the first approach with the tool to identify the candidate star formation site happens navigating a panoramic view of the Milky way. Therefore, the scientist is enabled to visually select the region of interest (with rectangular or circular selection as shown in Fig. \ref{fig1}), retrieving it automatically  (no additional technical details are needed to connect and get the data from the VLKB). The Milky Way panoramic view is obtained from the post processing of the Hi-GAL survey maps combining 70 $\mu$m (blue), 160 $\mu$m (green) and 250 $\mu$m (red) images using a novel multi-scale local stretching algorithm in false colour.

\begin{figure*}
\includegraphics[width=\textwidth]{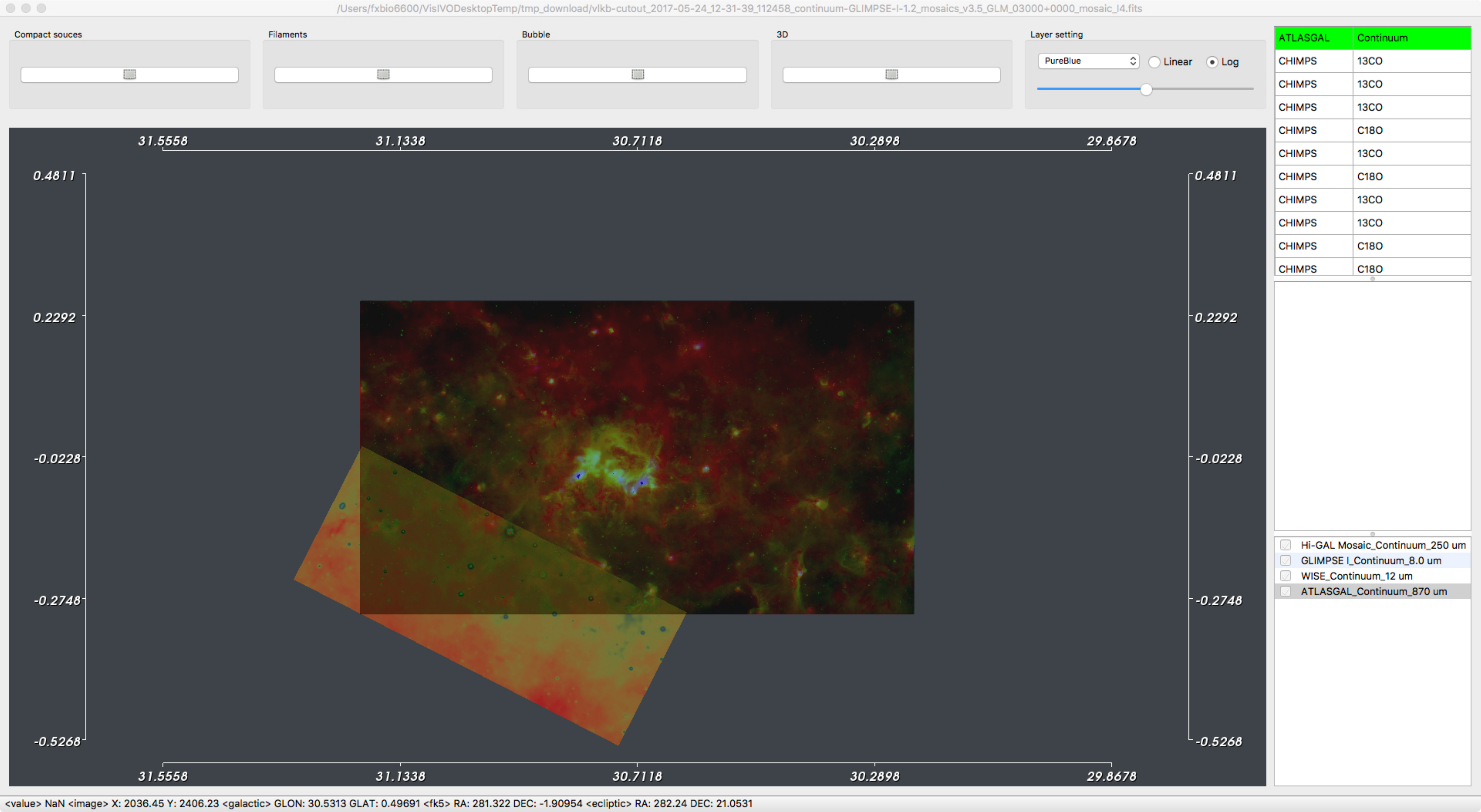}
\caption{Map region 2D visualization. Colour palette (with linear or logarithmic scale), contrast and saturation can be customized. Pixel values and coordinates are displayed at the bottom. A list of maps and data cubes from all the available surveys within the selected region is shown in the right panel.}
\label{fig2}
\end{figure*}

The FITS files retrieved within the user selected region, in the chosen survey and wavelength, downloaded from the Search \& Cutout service are displayed as shown in Fig. \ref{fig2} and, from this point on, it is possible to start the analysis operations. 

To better visualize and deeply analyze the astronomical object, the user can perform operations that are common in scientific visualization software: palette (lookup table) can be changed to map pixel values with different colours (with linear or logarithmic scale); the opacity, contrast and  saturation can be modified; layered images can be swapped. The status bar of the application shows the position of the mouse pointer over the image expressed as pixel (X-Y relative to the image), galactic, fk5 and ecliptic coordinates.

The access to the huge amount of information available within the user selected region is always at disposal of the scientist. Indeed VLVA, for each visualized image, presents a list of 2D maps and 3D datacubes stored in VLKB that come from the available surveys, automatically downloaded, aligned with the displayed \textit{base image} and over-imposed them as layer. This facilitates the correlation and comparison of information at different wavelengths and that comes from different instruments and space missions. 

\subsection{Visualization and analysis of Compact Sources }

In comparison to ground-based submillimeter	continuum surveys, the	data obtained by the Hi-GAL	project	with the \textit{Herschel} cameras do not suffer from the varying atmosphere, allowing therefore to recover the rich and highly structured large-scale emission from Galactic cirrus and extended clouds. Such variable and complex background severely	hinders the use of traditional methods to detect compact sources based on the thresholding of the intensity image widely used in the case of	large scale	millimeter and radio surveys from ground-based facilities, like the CORNISH or ATLASGAL where diffuse emission is filtered out either by the atmospheric variations or the instrumental transfer function. Therefore, a new a software tool,  called CuTEx, specialized to work in severe background conditions, such as in the Hi-GAL maps, has been developed \cite{molinari2011source}. In short, CuTEx algorithm highlights those regions computing the second derivative of the	map. All the ``clumps'' of pixels above	a defined threshold	are	analyzed	and the ones larger than a certain area are kept as candidate detections. The pixels of	the	large ``clumps'' are further analyzed to determine if there are more than one statistically significant local maximum of curvature, as expected in the case of very close sources. 

The compact sources extracted with the CuTEx package at the wavelength of 70 $\mu$m, 160 $\mu$m, 250 $\mu$m, 350 $\mu$m and 500 $\mu$m have been associated  basing on simple positional criteria for obtaining band merged catalogue. The basic step of this process is the association between two source lists corresponding to the detections obtained at two different wavebands $\lambda_2 < \lambda_1$. An association has been established between a source detected at $\lambda_2$ and one detected at $\lambda_1$ if the centroid of the former falls within the ellipse returned by CuTEx as the section at half maximum of the two-dimensional fit to the source profile. Cases of multiplicity, when more than one source fulfills the aforementioned criterion, are removed saving only the association of the closest counterpart at $\lambda_1$, and leaving the remaining ones unassociated. At this step, a list of entries composed by: i) associations between $\lambda_1$ and $\lambda_2$; ii) unassociated sources at $\lambda_1$; iii) unassociated sources at $\lambda_2$ is returned.
All these entries are considered for subsequent association with a third band $\lambda_3 < \lambda_2$, and so on.

For the Hi-GAL catalogue we start with $\lambda_1 =500 \mu$m and we carry on the associations with $\lambda_2 = 350 \mu$m, $\lambda_3 = 250 \mu$m and so on, according to the scheme described above. In this way, the band merged catalogue will contain sources having counterparts at all the five bands, or in fewer bands (adjacent or not), or even in a single band.
The resulting catalogue contains not only the photometric information, i.e. the fluxes at each band, but also the positional and morphological information about all the counterparts at all bands. For this reason, a generic source can be associated in the catalogue with up to five positions, each differing by a few arcseconds; to provide the user with a reference position, we define as master position the coordinates at the shortest available wavelength.

The data mining approach, named Q-FULLTREE (Quick Full TREe on Ellipse), has been applied to obtain band merged catalogues incorporating to the Hi-GAL catalogue also  source  catalogue  information  from  GLIMPSE,  GLIMPSE360,  WISE,  MIPSGAL,  ATLASGAL,  BGPS  and  Cornish  surveys. It is based on the same positional cross-match algorithm described above, but with a substantial difference in handling possible multiplicities.


In case there is more than one higher resolution source included in the ellipse, the value of the ellipse formula provides a by-product reliable and fast distance estimation of the object from the ellipse centre that can be used to assign a different score to all the candidate counterparts included in the same ellipse. Furthermore this scoring system takes implicitly into account the relationship among candidate sources, their bands and dimensions of ellipses.

VLVA represents an invaluable tool for the study of star cluster formation~\cite{molinari2016hi} offering visualization and analysis capabilities for compact sources, band-merged catalogues and Spectral Energy Distributions (SEDs).  


The scientist is enabled to perform all the operations fundamental to study the compact sources by selecting a sub-region on the map. Sources from the single-band and band-merged catalogues are automatically retrieved from the VLKB and displayed the  on top of the image. Compact sources are shown as ellipses with their relative FWHM and position angle using different colours for each wavelength. An example of compact sources visualization is shown in Fig. \ref{fig2}. 

\begin{figure}
\includegraphics[width=\columnwidth]{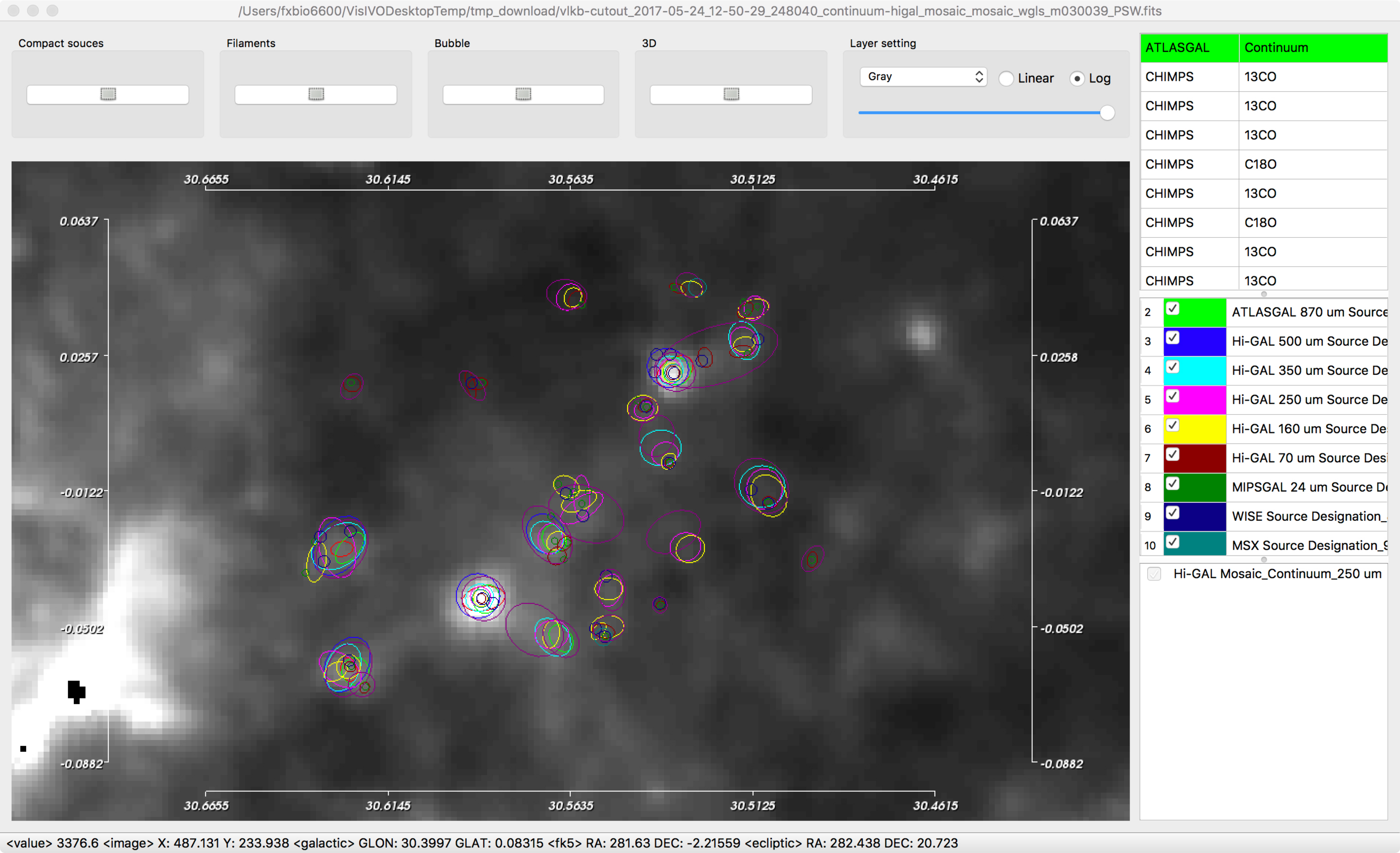}
\caption{Compact Sources visualization. Compact sources are shown in different colours on the image depending on the relative wavelength.}
\label{fig3}
\end{figure}

VLVA provides features to explore the physical parameters (stored into the VLKB) for each source, such as the SED (see Fig. 4). Moreover, 2D plots correlating any of two selected quantities can be easily produced. 

Selecting the retrieved band-merged compact sources, the SED analysis toolset can be started. This shows an interactive 2D plot of all the counterparts at different wavelengths, obtained by the Q-FULLTREE processing, that are related to the selected sources.

\begin{figure}
\includegraphics[width=\columnwidth]{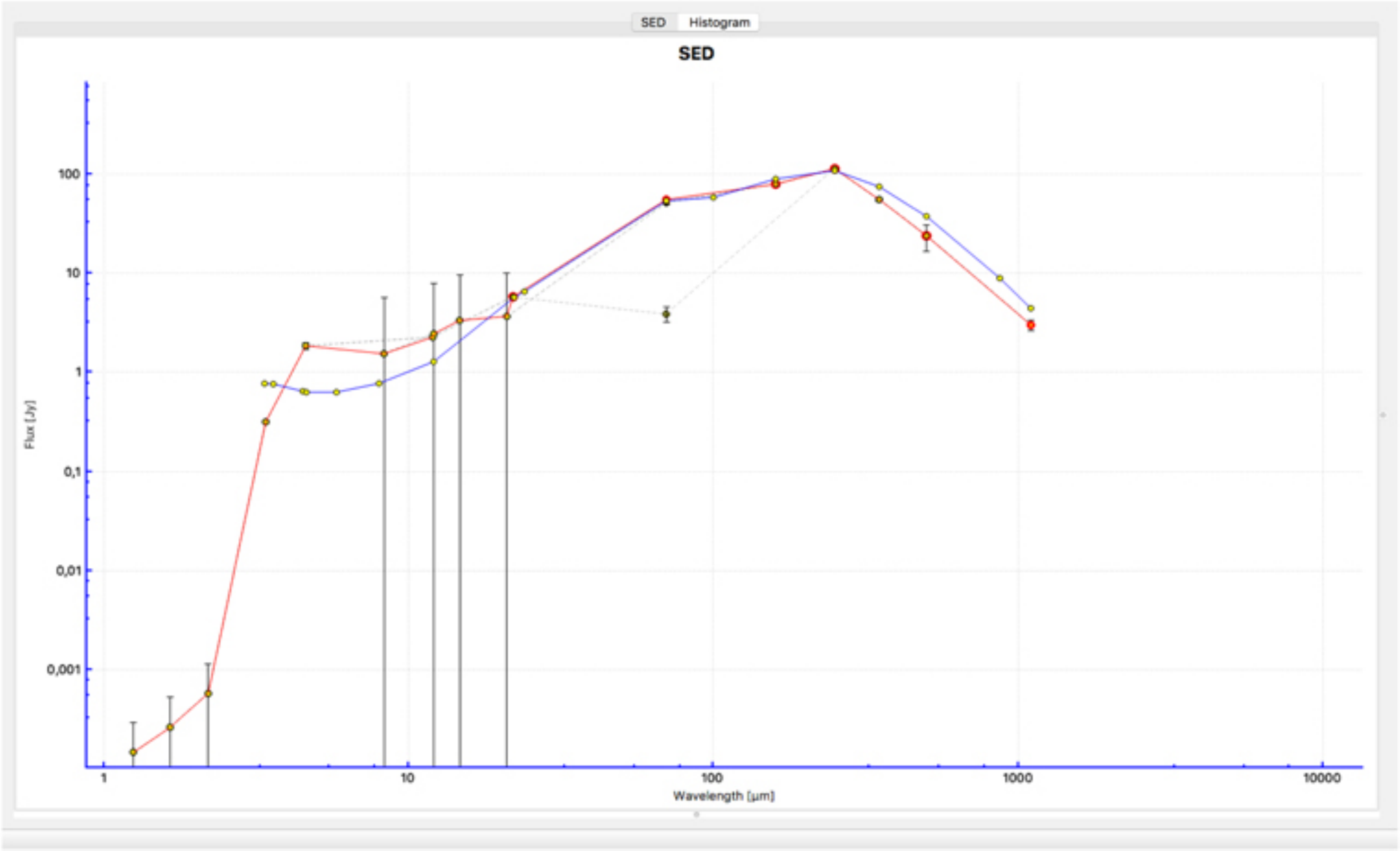}
\caption{SED analysis panel: the original SED (in dashed lines), the SED with cumulative fluxes and the theoretical models fit result.}
\label{fig4}
\end{figure}

In case the SED presents multiple associations the fluxes of counterparts can be summed obtaining the SED with a cumulative fluxes to be fitted (see Fig. \ref{fig4}). 

Three different kind of fitting operations are available: one for the fit with theoretical models \cite{robitaille2007interpreting} and two for the analytic grey-body fits (e.g. \cite{elia2016remarkable}). Fit operation is performed, by default, on all available wavelengths but a subset can be chosen. The computation is made through routines implemented using Interactive Data Language~\cite{IDL} (IDL) that access to the data stored within the VLKB. To avoid IDL dependencies on the VLVA, this fitting engine has been implemented as a remote service to be called by the visual analytics application in a transparent way for the user. The remote engine has been optimized to minimize data transfer and computing effort on the client side.
The logging information as well as the results of the computed SED fit are overplotted to the data points (see Fig. 4). For theoretical fitting operation, a list of best fitted SED models is displayed once a new fit is performed. Histogram of physical parameters of the SED fitting (e.g. mass and luminosity) can be easily calculated for the selected sources. The user can select/de-select from the resulting list the SED model to visualize it on the plot. 

To facilitate the scientific process and, in particular, the correlation between non-homogeneous information (e.g. SED, images, datacubes), the VLVA graphical interfaces are highly interactive. As an example, a selection on each SED point within the SED analysis toolset (Fig. \ref{fig4}) highlights the corresponding compact source in Fig. \ref{fig3}.

\subsection{Extended bubble-like sources}

To identify the physical agent responsible for triggering star formation in a cloud, it is necessary to search for a relationship between the nature and strength of the triggering agents, and the star formation rate (SFR) and star formation efficiency (SFE) as indicators of the intensity of star formation that they may have triggered. The new-generation Galactic Plane surveys offer now the possibility to investigate these relationships on a variety of spatial scales, from small bubbles around young HII regions mostly visible in the mid-to-far infrared and in the radio continuum, to arc-shaped portions of irregular HI superbubbles or SNR shell portions visible in the radio. Star forming clumps are analyzed close to these potential triggers to verify if SFR and SFE significantly differ from other locations in the Galaxy. Still such morphological evidence alone does not constitute unequivocal proof for a triggered star formation scenario. Compatibility between the average age of a triggered star-forming region and the dynamical age of the triggering agent is another additional check that can be done using the data at hand. A good starting point in this respect can be the bubble-like structures catalogued from the surveys (GLIMPSE and MIPSGAL) obtained by the NASA's Spitzer Space Telescope\cite{NASASST}. While GLIMPSE revealed more than 300 candidate bubbles mostly associated with hot young stars, nearly 90\% of the $\sim$400 colder dust bubbles revealed at 24 $\mu$m by MIPSGAL are still not identified even if there are strong hints indicating they represent a population of highly obscured, massive objects in later stages of their evolution. 

\begin{figure*}
\includegraphics[width=\textwidth]{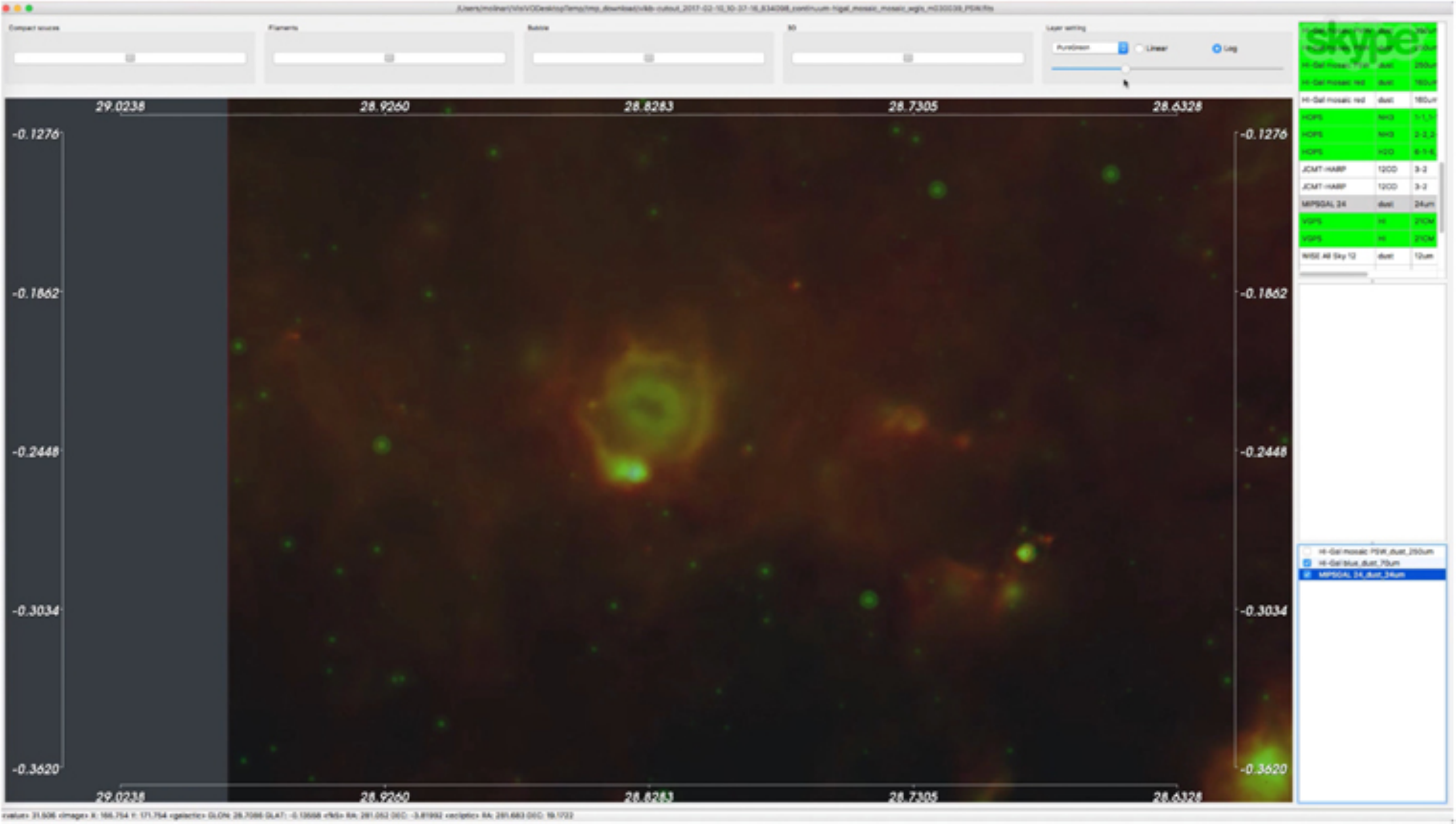}
\caption{Bubble-like structure visualization. An Hi-GAL map at 70 $\mu$m (with glow palette) superimposed on a 24 $\mu$m MIPSGAL image (with pure green palette).}
\label{fig5a}
\end{figure*}

The visual analytic application allows to investigate these regions by superimposing different maps from the available surveys. Fig. \ref{fig5a} shows a Hi-GAL map at 70 $\mu$m (glow palette) superimposed on a 24 $\mu$m MIPSGAL image (pure green palette). This allows to analyze dust emissions and dynamics: cold dust and large grains from the Hi-GAL map on the external ring of the bubble-like structure, and hot dust very fine grained from the MIPSGAL image in the internal part of the region.    

Bubbles can be checked against 3D molecular spectral datacubes (e.g. from maps of GRS surveys) to study the dust kinematics (as discussed in the next Section \ref{fil} for filamentary structures).  

\subsection{Filamentary structures}
\label{fil}

Filaments are considered key-structures required to	build the densities	necessary for star formation~\cite{schisano2014identification}. Whether driven by turbulence, magnetic field-mediated gravitational contraction, gravo-turbulent processes or shocks in large-scale Galactic flows, the transition from the diffuse and relatively low column density ISM into dense structures on the verge of star formation seems to pass through filamentary structures. Obtaining a complete census and an accurate physical characterization of dense filamentary structures over the entire extension of the Galactic Plane is a key milestone to identify the mechanisms responsible for assembling these structures and to compute the SFE.

\begin{figure*}
\includegraphics[width=\textwidth]{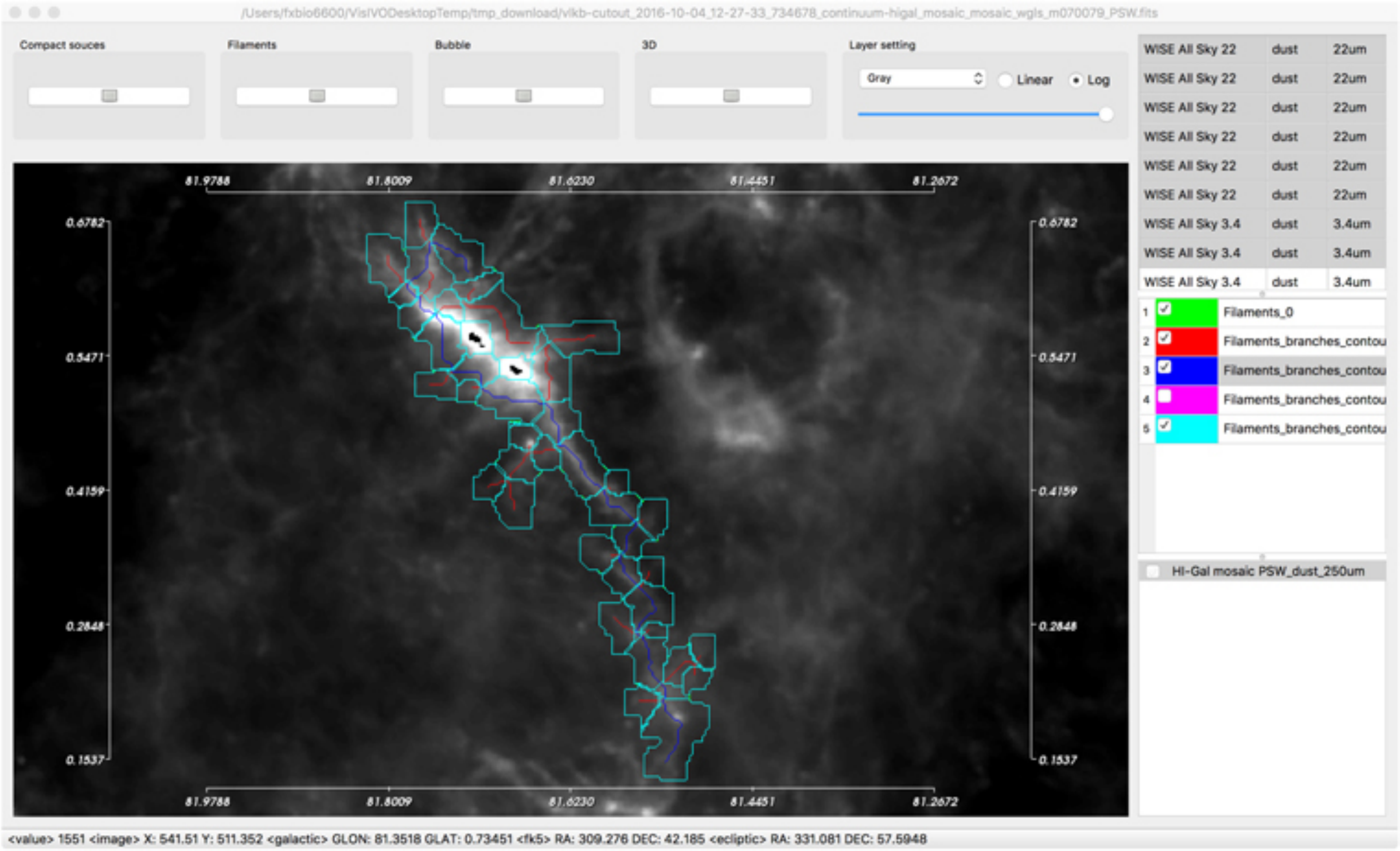}
\caption{Filamentary structures visualization using a set of 3 shapes, identifying filaments as the primary object, branches as the ``linear'' components within them and their spines, and nodes as the connection points of the various branch segments that compose a filament.}
\label{fig5}
\end{figure*}

Filaments may have highly variable lengths and widths, they may create complex web-like features and have variable orientation, but irrespectively of the pattern complexity they can be locally approximated as cylinder-like geometries and this may greatly facilitate the adoption of very specific algorithmic approaches. In particular the use of Hessian operators has been employed on the Hi-GAL survey to identify the candidate filaments stored in the VLKB~\cite{schisano2014identification}. 

\begin{figure*}
  \centering
  \subfigure[Datacube (containing velocity information)  visualization to aid filamentary structure inspection. The right panel shows a 2D representation of the velocity channel, selected with the cutting plane from the left panel. Isocontours can be changed by selecting the number of levels and the upper/lower bounds.]{\includegraphics[width=0.9\textwidth]{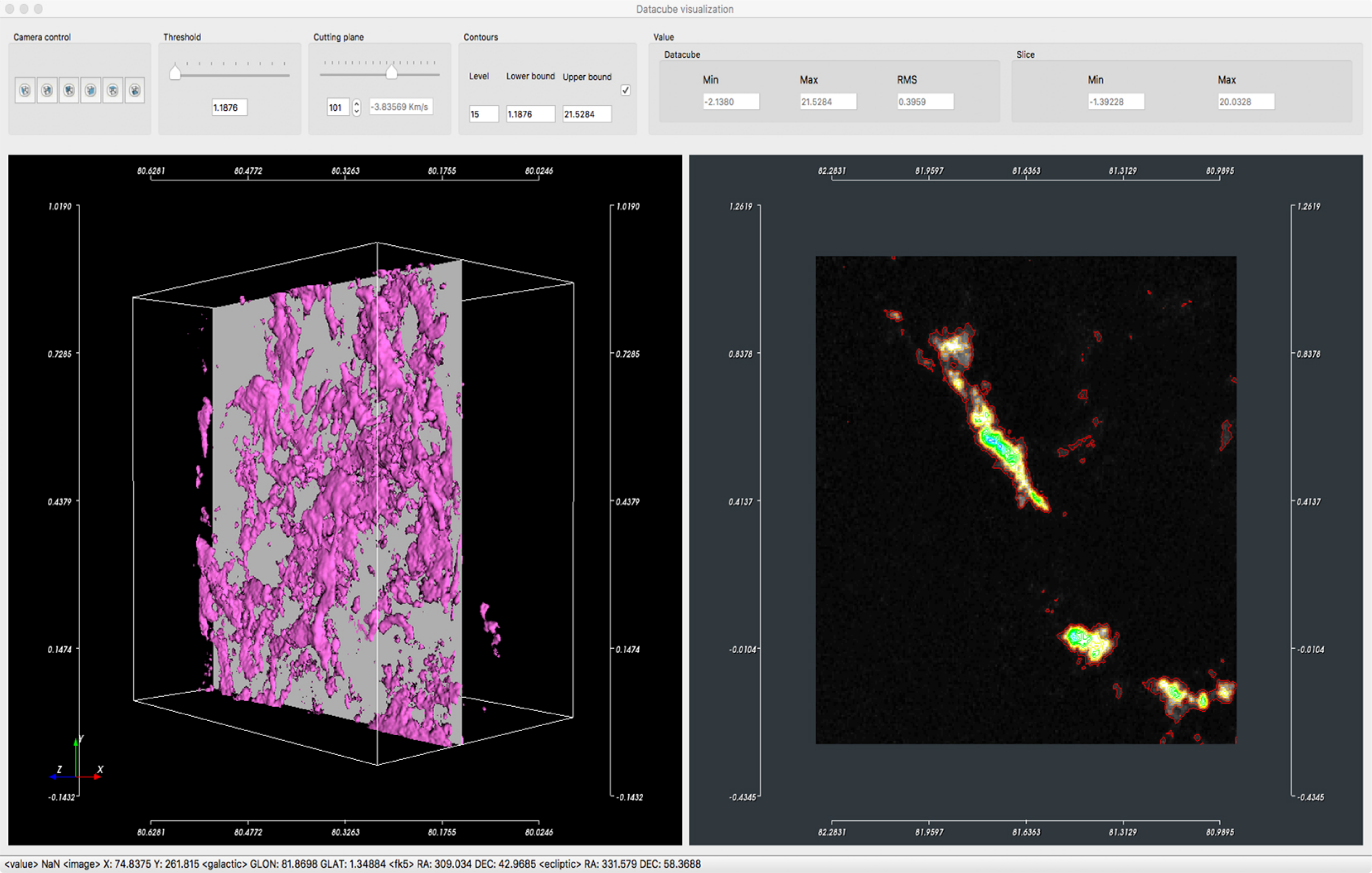}\label{datacube}}\quad
  \subfigure[Isocontours of the velocity channel shown in Fig. \ref{datacube} mapped on top of the filaments visualization.]{\includegraphics[width=0.9\textwidth]{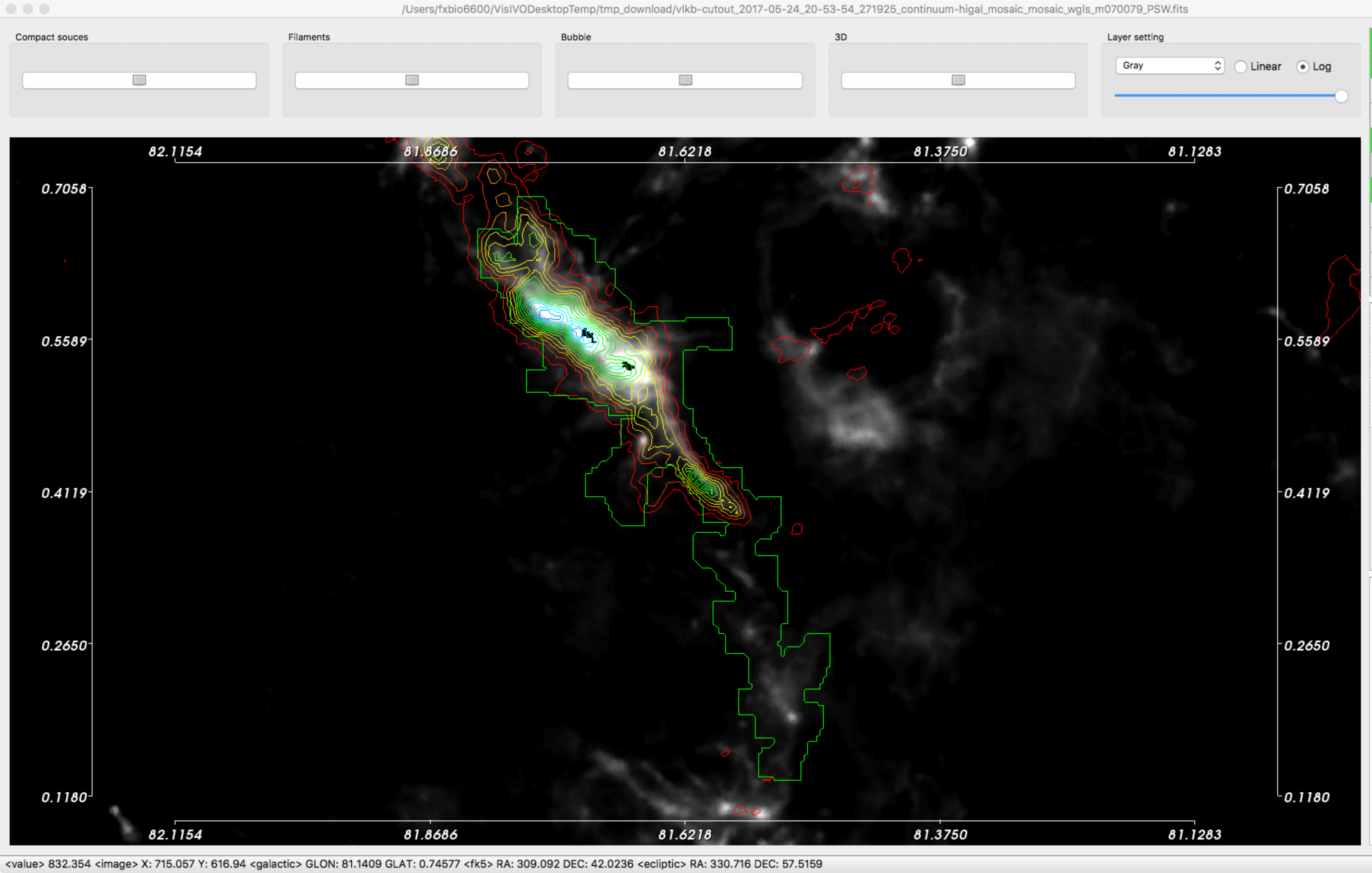}\label{fil_iso}}
  \caption{Datacube visualization and isocontours mapping on filamentary structures. This visualization is employed to verify that a candidate filament belongs to the same kinematic structure.}
  \label{fig6}
\end{figure*}

The candidate filaments are automatically downloaded from the VLKB and  displayed on top of the selected region on the base image as shown in Fig. \ref{fig5}. Filamentary structures are visualized using a set of 3 shapes, identifying filaments as the primary object, branches as the ``linear'' components within them and their spines, and nodes as the connection points of the various branch segments that compose a filament.

The visual inspection of filamentary structures can be a tricky task and can lead to misunderstanding results due to prospective effects of objects placed at different distances, therefore not physically connected. 
Thus, 2D visualization should be extended and validated with other information, in particular, those coming from the inspection of 3D molecular spectral datacubes to verify that a candidate filament belongs to the same kinematic structure (see Fig. \ref{fig6}). 

The molecular datacube, containing the velocity information, can be selected from the available surveys listed on the visualization window of the VLVA as shown in Fig. \ref{fig2}. The datacube 3D reconstruction is rendered using isosurfaces algorithm with chosen thresholds (see Fig. \ref{datacube}). The right panel in Fig. \ref{datacube} shows a 2D representation of the velocity channel, selected with the cutting plane from the left panel, and related isocontours that can be changed by selecting the number of levels and the upper/lower bounds.

Thanks to the VLVA interactivity between different visualization windows, the isocontours are also placed as additional layer on top of the map shown in Fig. \ref{fig5} to see the velocity mapping within the filamentary structures (see Fig. \ref{fil_iso}). 

\section{RELEVANT TECHNOLOGIES}

\begin{figure}
\centering
\includegraphics[width=0.5\textwidth]{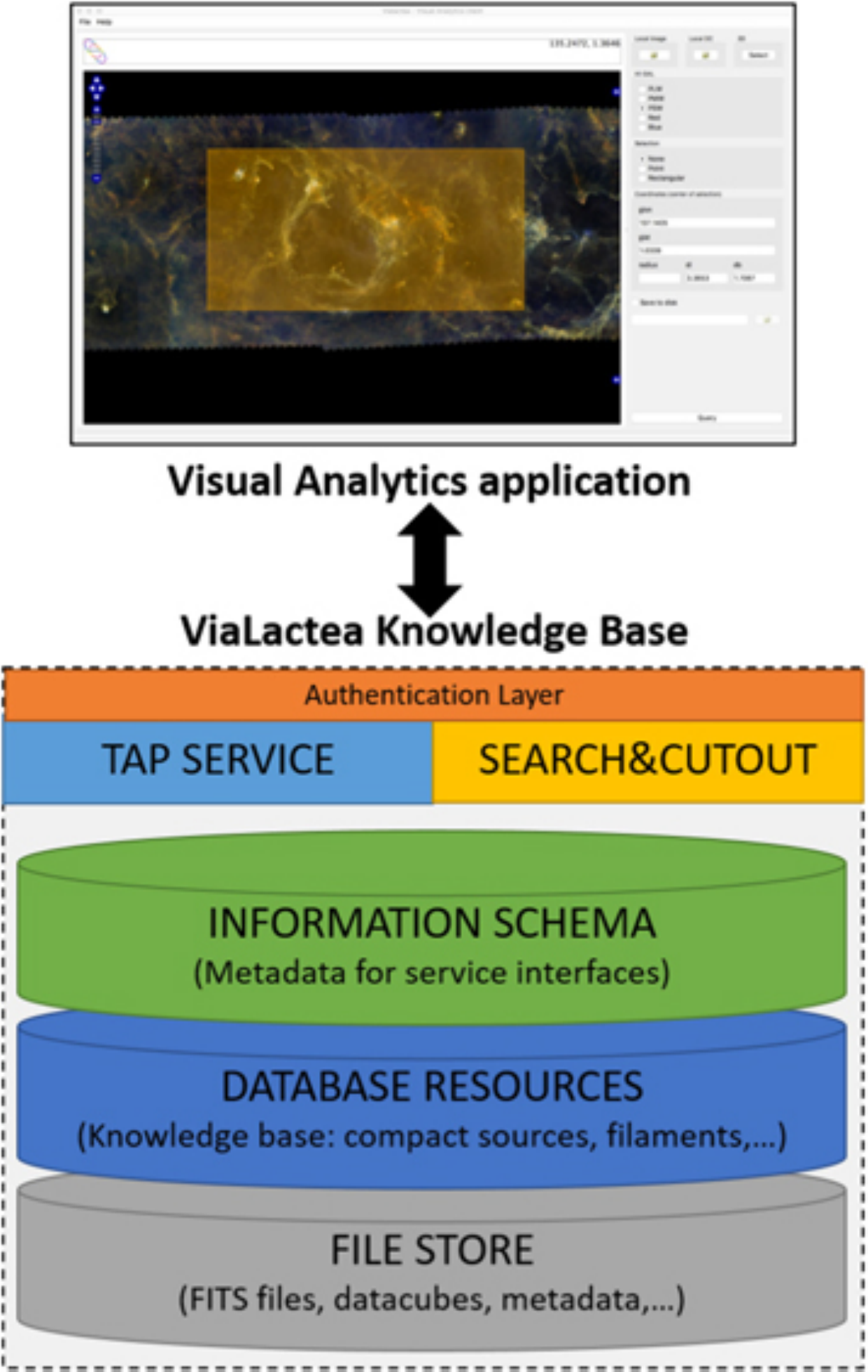}
\caption{Visual Analytics application interacting with the ViaLactea Knowledge Base (VLKB). VLKB exposes data and metadata trough the TAP service and the search \& cutout service.}
\label{arch}
\end{figure}

The VIALACTEA Visual Analytics is implemented as a cross-platform client-server application interacting (in a transparent way for the user) with the VLKB and the discovery and access interfaces as shown in Fig. \ref{arch}.

The core functionalities of VLVA are implemented in C++ and extended with Qt\cite{QT} for the graphical user interface. Qt enables a tight interactivity between different visualization windows thanks to its powerful events handling. 

The adopted rendering engine is the Visualization Toolkit\cite{VTK} (VTK). VTK is used worldwide in commercial applications, research and development, and is the basis of many advanced visualization applications such as ParaView\cite{ParaView}, VisIt\cite{Visit}, and VisIVO. VTK supports a wide variety of visualization algorithms including scalar, vector, tensor, texture, and volumetric methods; and advanced modeling techniques such as implicit modeling, polygon reduction, mesh smoothing, cutting and contouring. VTK includes an extensive information visualization framework and a suite of 3D interaction widgets; it supports parallel processing and integrates with various databases and GUI toolkits such as Qt. VTK is an open-source toolkit licensed under the BSD license. 

The VLVA overall data analysis and management is based on VisIVO Library \cite{becciani2012visivo} that makes direct use of VisIVO features mainly for the importing of tabular and 2D/3D images and filtering facilities (such as mathematical operations between physical parameters and computation of statistical information).

The access to VO-compatible databases and file store within the VLKB is performed using the Table Access Protocol\cite{TAP} (TAP) interfaces. The interoperability with the other VO-based tools (e.g. Aladin or TOPCAT) is guaranteed thanks to the implementation of the Simple Application Messaging Protocol\cite{SAMP} (SAMP).

As previously said, apart from the Hi-GAL surveys,  the other resources were retrieved from public repositories or released to the community retaining their private data policy. Therefore the access to the VLKB infrastructure has been secured via an authentication layer (currently based on HTTP basic access authentication). Alongside the data collections, a relational database (RDB) completes the VLKB resource content in terms of knowledge derived from the data and contains information related to: filaments and bubbles, compact sources and radio cubes.

All the data and metadata contents are discoverable
and accessible thanks to the interfaces that have been put on top of the database and storage system: a TAP service and a set of dedicated interfaces to consume stored datasets. While the TAP interface is useful for generic tabular access, it is less powerful to perform positional discovery and is not designed to provide direct access to data files. Therefore the search \& cutout service has been implemented to:

\begin{itemize}
\item identify the map images that overlap a region along a specific line of sight on the celestial sphere and described by a circle or a rectangular range around it
\item cutout the data files on positional constraints and also on the velocity axis (if 3D cubes are investigated) to allow for efficient data transport on the net and work only on the significant part of data
\item merge adjacent datasets into a single image/cube based on positional and velocity bounds.
\end{itemize}

\section{SUMMARY AND FUTURE WORKS}

We presented a visual analytics application that allows to combine 2D and 3D visualization with knowledge discovery and data mining methodologies accessing to new-generation surveys of the Galactic Plane to facilitate the studies of stars formation structures within our Galaxy. 

The application has been tailored to analyse compact sources with SEDs and extended sources such as bubbles and filaments. We demonstrated, through use cases, how the tool can easily support these analyses offering both qualitative and quantitative visualization capabilities.

Possible extensions to the current implementation that we are considering, include (1) enabling multiple users to interact with the same data set simultaneously, which will enable collaborative data analysis and visualization, (2) further optimizing the rendering and processing time using e.g. GPU technologies as in \cite{hassan2013tera}, and (3) investigating the use of European Open Science Cloud (EOSC) technologies for archive services related to the VLKB and intensive SED analysis employing the connection with the ViaLactea Science Gateway \cite{sciacca2017vialactea}.

\section*{Acknowledgements}
The research leading to these results has received funding from the European Union Seventh Framework Programme (FP7/2007-2013) under grant agreement no. 607380 (VIALACTEA) and from the European Commission’s Horizon 2020 research and innovation programme under the Grant Agreement no 739563 (EOSCPilot.eu).
\section*{References}
\bibliographystyle{plain}
\bibliography{biblio}

\end{document}